\documentclass[a4paper,11pt]{article} 
\usepackage[a4paper,top=2cm,bottom=2cm,left=3cm,right=3cm,marginparwidth=1.75cm]{geometry}

\usepackage[fleqn]{amsmath}
\usepackage{graphicx}
\usepackage{epstopdf}
\usepackage{fix-cm}
\usepackage{txfonts}
\usepackage{lipsum}
\usepackage{mathtools}
\usepackage{cuted}
\usepackage{multirow}
\usepackage{comment}
\usepackage{empheq}
\usepackage[titletoc,title]{appendix}

\usepackage{listings}
\usepackage{lineno}
\usepackage{mdwlist}
\usepackage{pennames}
\usepackage[british]{babel}
\usepackage{color}
\usepackage{subfig}
\usepackage{siunitx}
\usepackage{upgreek}
\usepackage{import}
\usepackage{amssymb}
\usepackage{epsfig}
\usepackage{cite}
\usepackage{hyperref}
\hypersetup{
colorlinks=true,
linkcolor=blue,
citecolor=blue,
urlcolor=blue,
}
\usepackage{titling}

\DeclareSIUnit \clight  {\textit{c}}
\DeclareSIUnit{\rad}{rad}

\RequirePackage[normalem]{ulem} 
\RequirePackage{color}\definecolor{RED}{rgb}{1,0,0}\definecolor{BLUE}{rgb}{0,0,1} 
\RequirePackage[normalem]{ulem} 
\RequirePackage{color}\definecolor{RED}{rgb}{1,0,0}\definecolor{BLUE}{rgb}{0,0,1} 

\DeclareGraphicsExtensions{.eps,.pdf,.jpeg,.jpg,.png}

\newcommand{\bea}{\begin{eqnarray}}
\newcommand{\eea}{\end{eqnarray}}
\newcommand{\be}{\begin{equation}}
\newcommand{\ee}{\end{equation}}
\newcommand{\fref}[1]{Fig.~\ref{#1}}
\newcommand{\Fref}[1]{Figure~\ref{#1}}
\newcommand{\sref}[1]{Sect.~\ref{#1}}

\newcommand*{\muonp}          {\ifmmode\mathrm{\muup^+}\else$\mathrm{\muup^+}$\fi}
\newcommand*{\muon}           {\ifmmode\mathrm{\muup}\else$\mathrm{\muup}$\fi}
\newcommand*{\tauon}          {\ifmmode\mathrm{\tauup}\else$\mathrm{\tauup}$\fi}

\newcommand*{\egamma}         {E_{\mathrm{\gammaup}}}
\newcommand*{\photon}         {\ifmmode{\gammaup}\else${\gammaup}$\fi}
\newcommand*{\positron}       {\ifmmode{\mathrm{e}^+}\else${\mathrm{e}^+}$\fi}
\newcommand*{\electron}       {\ifmmode{\mathrm{e}^-}\else${\mathrm{e}^-}$\fi}
\newcommand*{\eppair}       {\ifmmode{\mathrm{e}^\mathrm{e}^+} \else${\mathrm{e}^-}{\mathrm{e}^+}$\fi}
\newcommand*{\epositron}      {{E_\mathrm{e^+}}}

\newcommand*{\tegamma}        {{t_{\mathrm{e^+ \gammaup}}}}
\newcommand{\tg}{{\ifmmode t_{\gammaup_1\mathrm{e}^+}\else$t_{\gammaup_1\mathrm{e}^+}$\fi}}
\newcommand*{\tgg}{{\ifmmode t_{\gammaup\gammaup}\else$t_{\gammaup\gammaup}$\fi}}

\newcommand*{\thetae}         {{\theta_\mathrm{e^+}}}
\newcommand*{\phie}           {{\phi_\mathrm{e^+}}}

\newcommand{\meg}{\ifmmode{\muup \to e \gammaup}\else$\mathrm{\muup \to e \gammaup}$\fi}
\newcommand{\megp}{\ifmmode{\muup^+ \to \mathrm{e}^+ \gammaup}\else$\mathrm{\muup^+ \to e^+ \gammaup}$\fi}
\newcommand{\michel}{\ifmmode{\muup^+ \to e^+ \nuup\bar{\nuup}}\else$\mathrm{\muup^+ \to e^+ \nuup\bar{\nuup}}$\fi}
\newcommand{\radiative}{\ifmmode{\muup^+ \to \mathrm{e}^+\nuup\bar{\nuup}\gammaup} \else$\mathrm{\muup^+ \to e^+ \nuup\bar{\nuup}\gammaup}$\fi}
\newcommand{\conv}{\ifmmode{\muup^- \to e^-}\else$\mathrm{\muup^- \to e^-}$\fi}
\newcommand{\convN}{\ifmmode{\muup^-N \to e^-N}\else$\mathrm{\muup^-N \to e^-N}$\fi}
\newcommand{\mute}{\ifmmode{\muup \to 3e}\else $\mathrm{\muup \to 3e}$\fi}
\newcommand{\mutec}{\ifmmode{\muup^+ \to e^+e^+e^-}\else $\mathrm{\muup^+ \to e^+e^+e^-}$\fi}
\newcommand{\aif}{\ifmmode\mathrm{e}^+ \mathrm{e}^- \to \gammaup\gammaup \else$\mathrm{e}^+ \mathrm{e}^- \to \gammaup \gammaup$\fi}
\newcommand{\teg}{\ifmmode{\tauup \to e \gammaup} \else$\mathrm{\tauup \to e \gammaup}$\fi}
\newcommand{\tmg}{\ifmmode{\tauup \to \gammaup} \else$\mathrm{\tauup \to \muup \gammaup}$\fi}
\newcommand{\tmueg}{\ifmmode{\mathrm\tauup \to \ell \gammaup}\else$\mathrm{\tauup \to \ell \gammaup}$\fi}
\newcommand{\tautl}{\ifmmode{\mathrm\tauup \to 3\ell} \else$\mathrm\tauup \to 3\ell$\fi}

\newcommand*{\ypos}          {y_\mathrm{e^+}}
\newcommand*{\zpos}          {z_\mathrm{e^+}}

\newcommand*{\ugamma}         {u_{\gammaup}}
\newcommand*{\vgamma}         {v_{\gammaup}}
\newcommand*{\wgamma}         {w_{\gammaup}}

\newcommand*{\mathtentative}{}
\def\mathtentative#1#{\mathcoloraux{#1}}
\newcommand*{\mathcoloraux}[3]{%
  \protect\leavevmode
  \begingroup
    \color#1{#2}#3%
  \endgroup
}

\pretitle{\noindent\rule{\textwidth}{2pt}\begin{center}\LARGE\bfseries}
\posttitle{\end{center}\noindent\rule{\textwidth}{2pt}\vspace{-1cm}}

\begin{document}

\title{LETTER OF INTENT\\for a future \megp\ experiment\\at the High Intensity Muon Beam facility at PSI}


\date{}

\maketitle 

\pagenumbering{gobble}

\pagenumbering{arabic}

    
\begin{center}

{\bf \large THE STUDY GROUP FOR FUTURE \megp\ EXPERIMENTS}
\vspace{0.1cm}

Paolo Walter Cattaneo,$^{1}$${^*}$ Wataru Ootani,$^{2}$${^*}$ Francesco Renga,$^{3a}$${^*}$ Andr\'e Sch\"oning,$^{4}$${^*}$
Heiko Augustin,$^{4}$
Haris Avudaiyappan,$^{5}$
Sei Ban,$^{2}$
Paolo Beltrame,$^{6}$
Hicham Benmansour,$^{7ab}$
Daniela Bortoletto,$^{8}$
Alessandro Bravar,$^{9}$
Gianluca Cavoto,$^{3ab}$
Marco Chiappini,$^{7a}$
Alessandro Corvaglia,$^{18a}$
Giovanni Dal Maso,$^{10}$
Sacha Davidson,$^{11}$
Matteo De Gerone,$^{12}$
Lorenzo Ferrari Barusso,$^{12}$
Marco Francesconi,$^{13}$
Luca Galli,$^{7a}$
Giovanni Gallucci,$^{7}$
Flavio Gatti,$^{12}$
Helen Hayward,$^{6}$
Gavin Hesketh,$^{14}$
Malte Hildebrandt,$^{15}$
Fumihito Ikeda,$^{2}$
Fedor Ignatov,$^{6}$
Toshiyuki Iwamoto,$^{2}$
Tamasi Kar,$^{4}$
Marius K\"oppel,$^{16}$
Francesco Leonetti,$^{7ab}$
Weiyuan Li,$^{2}$
Ashley McDougall,$^{8}$
Satoshi Mihara,$^{17}$
Toshinori Mori,$^{2}$
Ljiljana Morvaj,$^{15}$
Donato Nicol\`o,$^{7ab}$
Hajime Nishiguchi,$^{17}$
Hiroyasu Ogawa,$^{2}$
Atsushi Oya,$^{2}$
Angela Papa,$^{7ab,}$$^{15}$
Marco Panareo,$^{18ab}$
Daniele Pasciuto,$^{3a}$
Davide Pinci,$^{3a}$
Richard Plackett,$^{8}$
Nikolaos Rompotis,$^6$
Massimo Rossella,$^{1}$
Thomas Rudzki,$^{4}$
Rei Sakakibara,$^{2}$
Susanna Scarpellini,$^{3ab}$
Taikan Suehara,$^{2}$
Hiromu Suzuki,$^{19}$
Masato Takahashi,$^{19}$
Michele Tammaro,$^{20}$
Gianfranco Tassielli,$^{18ac}$
Yusuke Uchiyama,$^{17}$
Ryusei Umakoshi,$^{2}$
Antoine Venturini,$^{7ab}$
Luigi Vigani,$^{4}$
Cecilia Voena,$^{3ab}$
Joost Vossebeld,$^{6}$
Rainer Wallny,$^{16}$
Kensuke Yamamoto,$^{2}$
Yuji Yamazaki$^{19}$
\vspace{0.5cm}
\normalsize
\end{center}
\noindent
$^1$ INFN Sezione di Pavia, Via Bassi 6, 27100 Pavia, Italy \\
$^2$ ICEPP, The University of Tokyo, 7-3-1 Hongo, Bunkyo-ku, Tokyo 113-0033, Japan\\
$^3$ a) INFN Sezione di Roma, Piazzale A. Moro, 00185 Rome, Italy \\
\phantom{$^3$} b) Dipartimento di Fisica dell’Università “Sapienza”, Piazzale A. Moro, 00185 Rome, Italy \\
$^4$ Physikalisches Institut, Universit\"at Heidelberg, Im Neuenheimer Feld 226, 69120 Heidelberg, Germany\\
$^{5}$ Institut f\"ur Kernphysik und Exzellenzcluster PRISMA+, 
Johannes Gutenberg-Universit\"at Mainz, Johann-Joachim-Becher-Weg 45, 55128 Mainz, Germany \\
$^6$ Oliver Lodge Laboratory, University of Liverpool, Oxford St., Liverpool L69 7TE, UK\\
$^7$ a) INFN Sezione di Pisa, Largo B.~Pontecorvo~3, 56127 Pisa, Italy \\
\phantom{$^7$} b) Dipartimento di Fisica dell'Universit\`a, Largo B.~Pontecorvo~3, 56127 Pisa, Italy \\
$^{8}$ Department of Physics, University of Oxford, Denys Wilkinson Building, Keble Road, Oxford OX1 3RH, United Kingdom \\
$^{9}$ D\'epartement de physique nucl\'eaire et corpusculaire,
Universit\'e de Gen\`eve, 24, quai Ernest-Ansermet, 1211 Gen\'eve 4, Switzerland \\ 
$^{10}$ CERN, Esplanade des Particules 1, 1217 Meyrin, Switzerland\\
$^{11}$ Institut de Physique des deux Infinis de Lyon, 4 rue Enrico Fermi, 69622 Villeurbanne, France \\
$^{12}$ a) INFN Sezione di Genova, Via Dodecaneso 33, 16146 Genova, Italy \\
\phantom{$^12$} b) Dipartimento di Fisica dell’Universit\'a, Via Dodecaneso 33, 16146 Genova, Italy \\
$^{13}$ INFN Sezione di Napoli, Via Cintia, 80126 Napoli, Italy \\
$^{14}$ Department of Physics and Astronomy, University College London, Gower Street, London WC1E 6BT, United Kingdom \\
$^{15}$ PSI Center for Neutron and Muon Sciences, 
Forschungsstrasse 111, 5232 Villigen PSI, Switzerland \\
$^{16}$ Institute for Particle Physics and Astrophysics, Eidgenössische Technische Hochschule Zürich, Otto-Stern-Weg 5, 8093 Zürich, Switzerland\\
$^{17}$ KEK, High Energy Accelerator Research Organization, 1-1 Oho, Tsukuba, Ibaraki 305-0801, Japan\\
$^{18}$ a) INFN Sezione di Lecce, Via per Arnesano, 73100 Lecce, Italy \\
\phantom{$^{18}$} b) Dipartimento di Matematica e Fisica
dell'Universit\'a del Salento, Via per Arnesano, 73100 Lecce, Italy \\
\phantom{$^{18}$} c) Departement of Engineering and Science,
Mercatorum University, Piazza Mattei, 10 , 00186 Roma \\
$^{19}$ Kobe University, 1-1 Rokkodai-cho, Nada-ku, Kobe, Hyogo 657-8501, Japan\\
$^{20}$ Galileo Galilei Institute for Theoretical Physics, Largo Enrico Fermi 2, 50125 Firenze, Italy\\

\vspace{0.5cm}
\noindent
${^*}$ Contact persons: \\
\href{mailto:paolo.cattaneo@pv.infn.it}{paolo.cattaneo@pv.infn.it} \\\href{mailto:wataru@icepp.s.u-tokyo.ac.jp}{wataru@icepp.s.u-tokyo.ac.jp} \\
\href{mailto:francesco.renga@roma1.infn.it}{francesco.renga@roma1.infn.it} \\ 
\href{mailto:schoning@physi.uni-heidelberg.de}{schoning@physi.uni-heidelberg.de}


\clearpage

\begin{abstract}
\large
Searches for charged lepton flavor violation in the muon sector stand out among the most sensitive and clean probes for physics beyond the Standard Model. Currently, \megp\ experiments provide the best constraints in this field for a wide range of models while, in the coming years, new experiments investigating the processes of \mutec\ and $\muup \to \mathrm{e}$ conversion in the nuclear field are anticipated to reach comparable or higher sensitivities. 
The High-Intensity Muon Beam (HIMB) facility at PSI, which is expected to deliver muon beam intensities up to two orders of magnitude higher than the existing beam lines, offers a unique opportunity to significantly enhance the sensitivity of \megp\ searches. The discovery potential could be substantially boosted and a sensitivity comparable to that of all the other projects could be reestablished, which is essential for discriminating among competing new-physics scenarios should an observation occur in any of the channels.
In this document, we express our interest in developing a \megp\ experimental program at HIMB, with the goal of improving, within the next decade, the sensitivity of the \megp\ search by more than one order of magnitude relative to the expected final result of the current leading experiment, MEG II. This effort would ensure that PSI retains its leadership in this field.

\end{abstract}

\newpage

\tableofcontents




\section{Introduction}

In the standard model (SM) of particle physics, charged lepton flavour-violating 
(cLFV) processes  are basically forbidden with only extremely small 
branching ratios (\num{\sim e{-54}}~\cite{calibbi_2018}) when
accounting for non-zero neutrino mass differences and mixing angles. 
However, being the conservation of the lepton flavor an accidental symmetry in the SM, several New Physics (NP) extensions predict cLFV decays at measurable rates, and the channel \megp\ is particularly sensitive. Thus, such decays serve as an excellent probe for NP, due to their high sensitivity and the absence of SM contributions, making any potential positive signal unequivocal evidence for physics beyond the SM. Reviews of the theoretical expectations and experimental status are provided in \cite{calibbi_2018,Mihara:2013zna}.

The MEG~II experiment set the best limit on the branching ratio of the \megp\ decay, $\mathcal{B}(\megp) < 1.5 \times 10^{-13}$ at \SI{90}{\percent} confidence level~\cite{MEGII:2025gzr}, providing the best cLFV constraints in the muon sector for a wide selection of NP models. MEG~II is expected to continue taking data in 2026, reaching a sensitivity of \num{6e-14}. 
In the meanwhile the Mu3e \cite{mu3edesign}, Mu2e \cite{mu2etdr} and COMET \cite{COMET:2018auw} collaborations are finalizing the construction of their detectors to search for the \mutec\ decay and the \convN\ conversion in the Coulomb field of nuclei, with sensitivities to NP comparable to 
or better than the MEG~II one. 
However, given the possibly different NP contributions to these processes (at tree level, \megp\ requires a dipole-like vertex, while the others can also go through an effective four-fermion contact interaction), these searches are complementary. 
If an observation is made in one of these experiments, only the comparison with the results of the others allows to discriminate between different NP models. This is only possible if their sensitivities (in terms of NP mass scale) are at comparable levels. 

\begin{figure}[!htb]
\centering
\includegraphics[width=0.6\textwidth,angle=0] {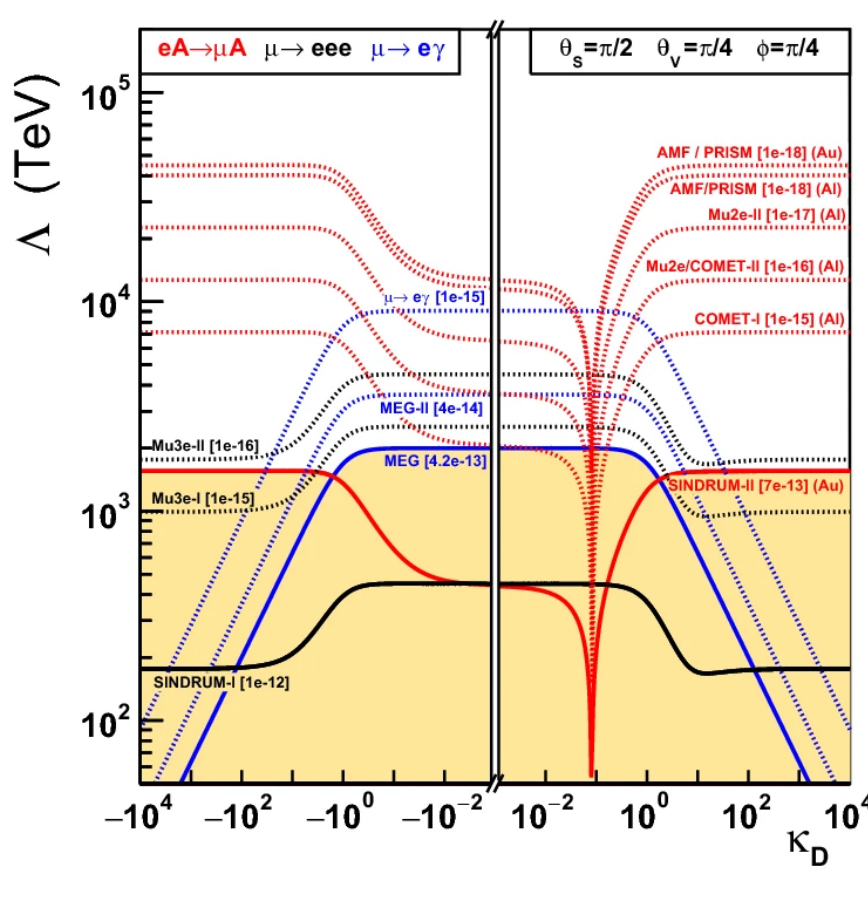}
  \caption{NP scale sensitivity of the different LFV muon channels as a function of a parameter $\kappa_D$ which tunes the relative magnitude of dipole and four-fermions coefficients in an effective field theory approach ($\kappa_D = 0$ for a pure dipole interaction) from \cite{Davidson:2022nnl}}
 \label{fig:theory}
\end{figure}

\Fref{fig:theory}, from~\cite{Davidson:2022nnl}, shows the NP scale sensitivity of the different channels as a function of a parameter $\kappa_D$ which tunes the relative magnitude of dipole and four-fermions coefficients in an effective field theory approach ($\kappa_D = 0$ for a pure dipole interaction), for a specific choice of other relevant parameters. Reaching with \megp\ a sensitivity competitive with the already approved experiments in the dipole-dominated region would be already possible with a moderate improvement of the MEG~II limit, i.e. reaching the $10^{-14}$ level on the branching ratio, while going down to a few $10^{-15}$ will be necessary to compete with the ultimate \convN\ experiments that have been proposed.

With this document, we express our interest to the development of a future \megp\ experiment at the Paul Scherrer Institute (PSI). A new detector concept is presented. 
With a staged approach, it is expected to improve the MEG~II sensitivity down to $\sim 10^{-14}$ in Phase-I at the existing HIPA facility
and reach a few $10^{-15}$ branching ratio sensitivity in Phase-II when exploiting the high intensity muon rates (HIMB) that 
will be available after the upgrade of the existing muon beam facilities.

With an appropriate design, this project could be easily complemented to extend its physics case to the search of a wide family of rare muon decays, including the aforementioned \mutec\ and decays with elusive or long-lived non-standard particles, following the paths already explored for MEG II~\cite{Grandoni:2025pcs} and Mu3e~\cite{Knapen:2024fvh}.

\section{Current experimental status}

\begin{figure}[!htb]
\centering
\includegraphics[width=0.7\textwidth,angle=0] {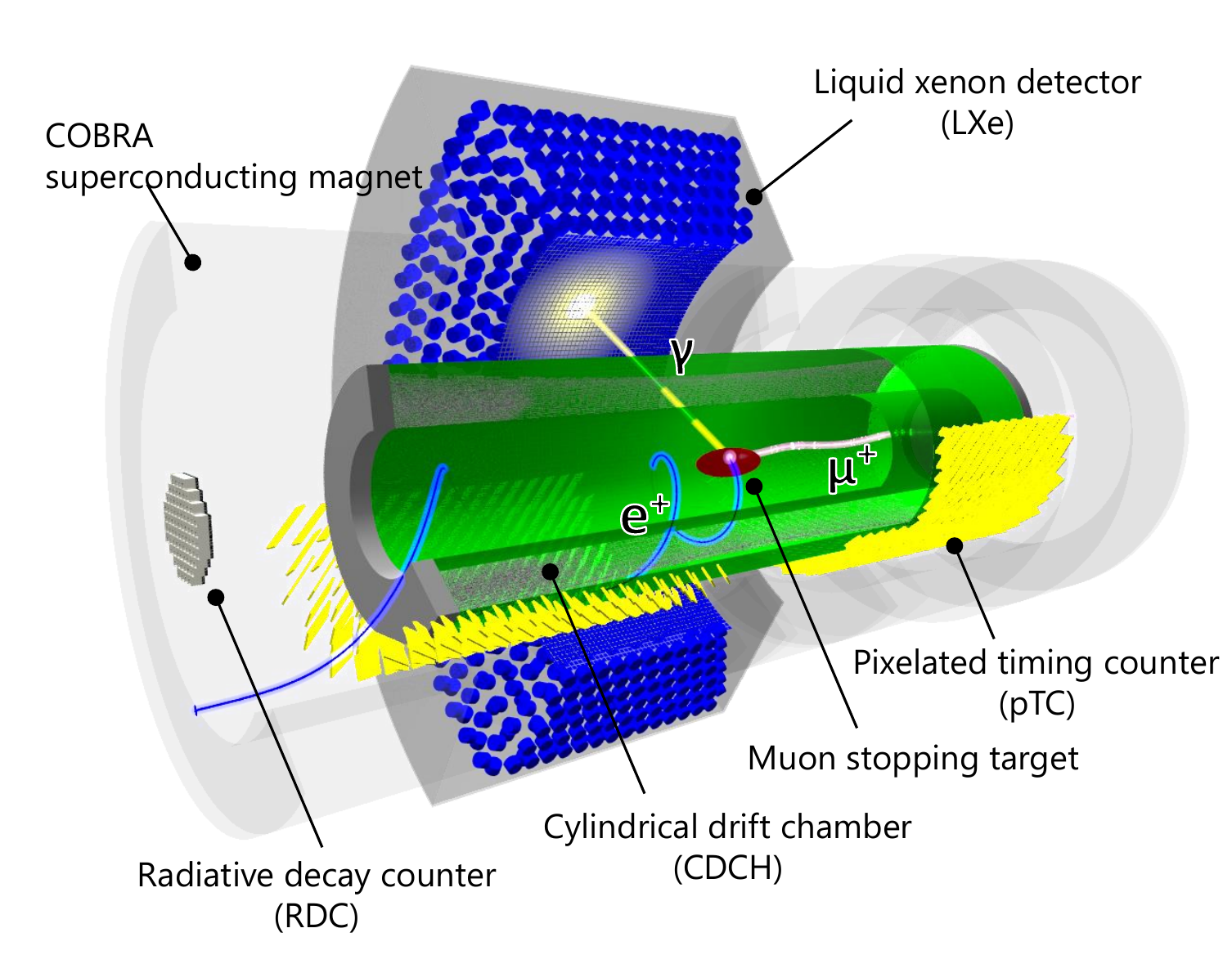}
  \caption{A sketch of the MEG~II experiment with a simulated event.}
 \label{meg2det}
\end{figure}

The MEG~II experiment, currently being conducted at the PSI, is the most sensitive \megp\ search to date.
The extremely high sensitivity of MEG~II relies on the high intensity continuous muon beams available at PSI and the innovative detectors identifying photons and positrons.
The MEG~II experiment was upgraded from MEG, retaining the experimental concept but with significantly improved detector performance (\fref{meg2det})\cite{baldini_2018,megII-det}.

The decays of positive muons stopped in a thin target at a rate of 
$(3-5)\times 10^7 \upmu^+/\mathrm{s}$ 
are measured by the positron and photon detectors surrounding the target with a solid angle coverage of 10\%. 
The positron detector is a magnetic spectrometer with a gradient magnetic field designed to quickly sweep away low-momentum Michel positrons 
from the tracking volume.       
The positrons are precisely tracked by a low-mass single-volume cylindrical drift chamber 
with a stereo wire configuration to form high-granularity drift cells.  
The positron timing is measured by a segmented timing detector composed of 
512 fast scintillator tiles read out by Silicon PhotoMultipliers (SiPMs).
The signal positron passes through multiple scintillator tiles in the detector. An excellent timing resolution is achieved by averaging the times measured by the multiple tiles.
The energy, timing, and impinging position of the photon are measured by liquid xenon (LXe) calorimeter with a single-volume 900L LXe surrounded by VUV-sensitive photosensors; 
4092 MPPCs ($12\times 12\,\mathrm{mm}^2$) on the front face for fine photon imaging and 668 PMTs (2-inch) on the other faces.
A new detector, Radiative Decay Counter (RDC), 
was introduced in MEG~II to further suppress the photon backgrounds 
from the radiative muon decays (RMD) 
by identifying low momentum Michel positrons associated 
to RMD photons around the signal photon energy.
The RDC, composed of plastic scintillators for timing and LYSO crystals for energy measurements, is placed on the beam axis only at the downstream end.

\section{Beam requirements}

Experiments dedicated to the search for rare decays, such as MEG/MEG~II and Mu3e, require the collection 
of high statistics to achieve their sensitivity goals. The current proposal requires two different beam configurations: a medium muon decay rate up to $R_\mu=$\SI{2e8}{\upmu^+/\second} in a first phase, which can be achieved with the current High Intensity Proton Accelerator (HIPA) facility in one of the existing experimental areas, followed in a second phase by a high muon decay rate delivered by 
the High Intensity Muon Beams (HIMB) project \cite{CDR},
capable of delivering a high rate up to $R_\mu=$\SI{e10}{\upmu^+/\second}.

\begin{figure}[!h]
        \centering
        \includegraphics[width = 0.7\textwidth]{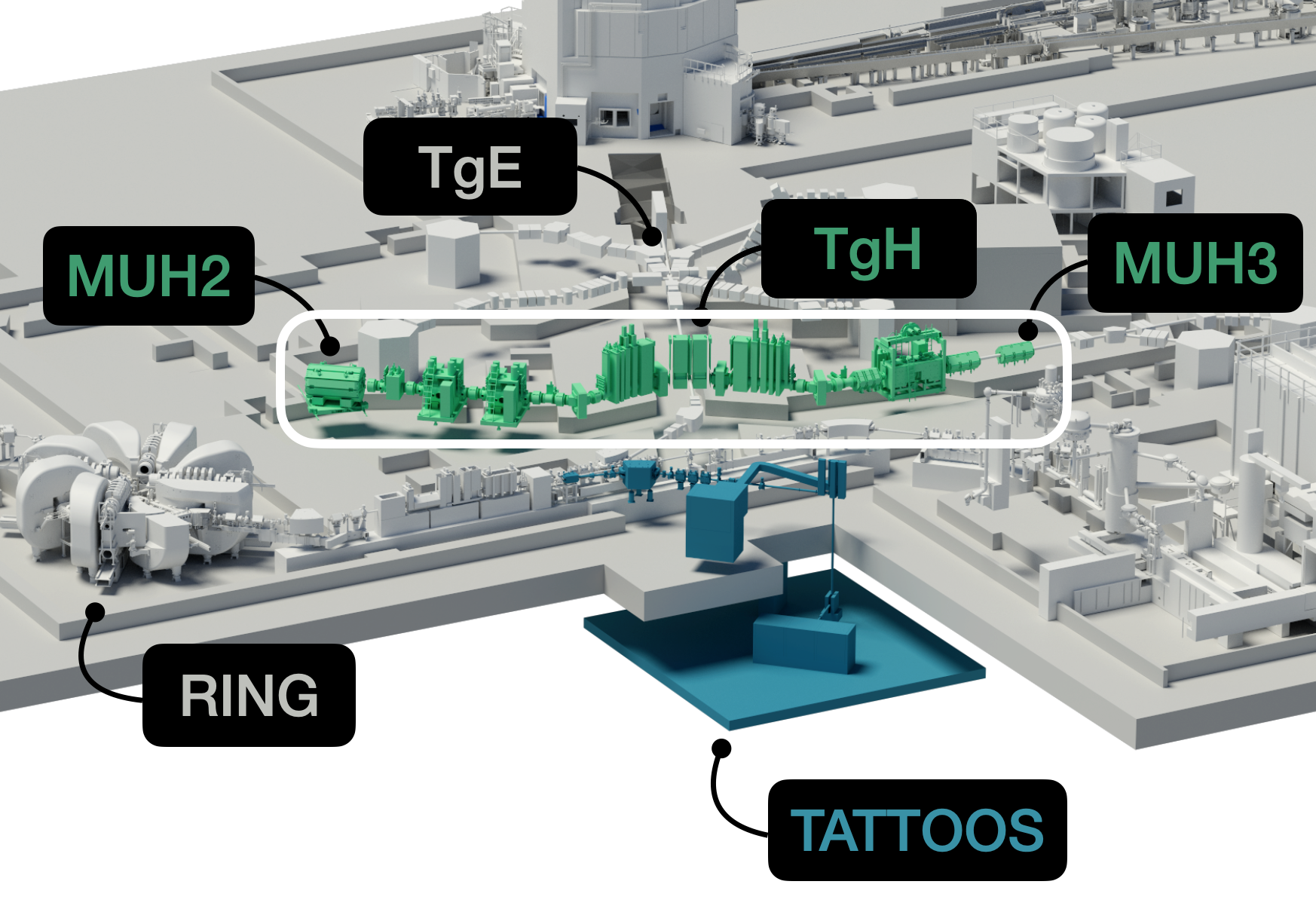}
        \caption{\textit{Overview of the HIPA accelerator complex including the HIMB beam lines and the new isotope production target station TATTOOS. The new target station TgH will replace the current TgM and the beam lines will replace $\uppi$M1 and $\uppi$M3. Adapted from \cite{CDR}.}}
	\label{f:newHIPA}
\end{figure}

\Fref{f:newHIPA} shows the layout of the HIPA facility including the HIMB beam lines, that will be installed at the HIPA shutdown during 2028-2029.

The current designs 
will allow to deliver up to \SI{\sim 1.10e10}{\upmu^+/\second} with a beam spot size on target of $\sigma_{x,y}=\SI{35}{\milli\meter}\times\SI{35}{\milli\meter}$ and a positron contamination lower than \SI{5}{\percent}.

In addition, the beamline is designed to deliver particles with momenta up to \SI{80}{\mega\electronvolt/c} and with the possibility to switch to electron, positron and charged pion beams for further experimental applications \cite{scienceCase, thesisGDM}. 

\section{Experimental concept}
\label{sec:expcon}

As in MEG/MEG~II, the experimental concept is based on reconstructing photons and positrons from \muonp s decaying at rest in a thin stopping target. 

The single event sensitivity (SES) is defined as 
the \megp\ branching ratio for which the expected number of 
detected decays is equal to one:
\begin{equation}
SES = \frac{1}{R_{\mu} \cdot T \cdot (\Omega/4\pi)} \times \frac{1}{\epsilon_e \cdot \epsilon_\gamma \cdot \epsilon_{trg}}
\label{eq:ses}
\end{equation}
where $R_\mu$ is the muon stop rate on target, $T$ the total time of data taking, $\Omega$ is 
the solid angle covered by the experiment, $\epsilon_e$ is the signal positron detection efficiency, $\epsilon_\gamma$ is the signal photon detection efficiency, $\epsilon_{trg}$ 
is the trigger efficiency.


The background is due to two main sources: the Radiative Muon Decays (RMD) or the accidental 
background with $e^+$ and $\gammaup$ coming from different $\muup$s. At high rate the latter
dominates and is the only one considered in the following.

The two-body kinematics is exploited to discriminate signal from backgrounds. It requires excellent resolutions on photon and positron energies, their relative direction and their relative emission time. An approximate estimation of the effective branching ratio of the accidental background is obtained from the formula in \cite{PhysRevD.65.112002}:
\begin{equation}
B_{acc} = R_{\muup} \cdot \delta t_{e \gamma} \cdot \delta x \cdot (\frac{\delta y}{15})^2 \cdot (\frac{\delta \theta_{e\gamma}}{2})^2 \cdot f(\theta_\gamma)
\label{eq:bck}
\end{equation}
where $\delta t_{e \gamma}$ is the uncertainty on the relative time between the positron and the $\gamma$-ray, $\delta x$ is the relative uncertainty
on the positron energy, $\delta y$ is the 
relative uncertainty on the $\gamma$-ray energy, 
$\delta \theta_{e\gamma}$ is the uncertainty on the 
relative angle between the positron and the $\gamma$-ray directions
and the uncertainties are to be understood as FWHM; 
$f(\theta_\gamma)$ is the function defining the background reduction factor due to the measurement of the $\gamma$-ray angle. As a consequence of the dependence of $B_{acc}$ on $R_{\muup}$ (which implies a dependence of the background rate on $R_{\muup}^2$), if a non-negligible background yield is expected over the foreseen lifespan of the experiment, an increase of the beam rate cannot produce any significant improvement of the sensitivity, which would be only possible with an improvement of the resolutions.

The best resolutions on \positron\ energy and direction are offered by magnetic spectrometers. Given the relatively low momentum of signal positrons (\SI{52.8}{\mega\electronvolt/c}), 
Coulomb scattering in their interaction with the detector and passive material plays a critical role in determining the achievable resolutions in momentum and direction. 
At the state of the art, gaseous detectors with helium-based mixtures guarantee the best compromise of single-hit resolution and material budget. 
Nevertheless, drift chambers have limited rate capabilities due to poor granularity and aging effects, while conventional time projection chambers (TPCs) cannot provide the necessary single-hit resolutions when operated with helium~\cite{Baldini:2013ke_back}, 
although alternative geometries could be considered (as, for instance, TPCs with a radial drift field).
For this reason, while we pursue the goal of using drift chambers as the baseline solution for medium $R_\mu$, silicon pixel detectors will be the choice for high $R_\mu$.
For the former case, we will rely on the experience gained with MEG~II that should be readily applicable.
For the latter case, R\&D studies are being performed in the context of the Mu3e Phase~II upgrade project~\cite{scienceCase} aiming to further reduce the material of ultra-thin tracking detectors which are based on High Voltage Monolithic Active Pixel Sensors (HVMAPS)~\cite{Peric:2007zz} and have \SI{0.1}{\percent} radiation length or less. 

On the photon side, the performance of the LXe calorimeter of MEG/MEG~II largely surpassed all previous experiments.
However, there is no clear path to further improve the photon energy resolution in this kind of detector, limiting the profitable beam rate and the sensitivity to the level of the MEG~II ones. 
Innovative crystals like LYSO could provide better performance, and R\&D activities are ongoing on small prototypes~\cite{PapaLYSO}. 
In this case, a limiting factor could be the technological capability of growing large crystals, and their cost. 
Hence, although this calorimetric approach can be an interesting option (see Sec.~\ref{sec:schedule}), we are considering a different approach as a baseline: a thin layer of dense material would be used to convert photons into $\positron\electron$ pairs, subsequently tracked in a magnetic spectrometer. 
In principle, an excellent momentum resolution can be achieved on the $\positron\electron$ pair. 
However, if a passive conversion layer is used, the energy loss fluctuations in the converter itself limit the energy resolution~\cite{Cavoto2018}. 
It calls for a very thin conversion layer, resulting in poor conversion efficiency. 
This limitation can be partially overcome, adopting thin LYSO crystals readout by SiPM as active converters, where the ionization energy loss can be measured and used to correct the energy reconstructed in the gaseous \positron\electron\ spectrometer\cite{Sakakibara:2025yhk,Ban:2025sev}.
The unrecoverable energy loss will be limited to the radiation losses in the layer that will be undetected, and multiple layers of active converter and tracker can be stacked to further increase the efficiency. However, this solution would still require a higher $R_\mu$ to compensate for the low detection efficiency and surpass the calorimetric approach.

\subsection{Magnetic field configuration}

\subsubsection{Single solenoid magnetic structure}

\begin{figure}[]
\centering
\includegraphics[width=\textwidth,angle=0] {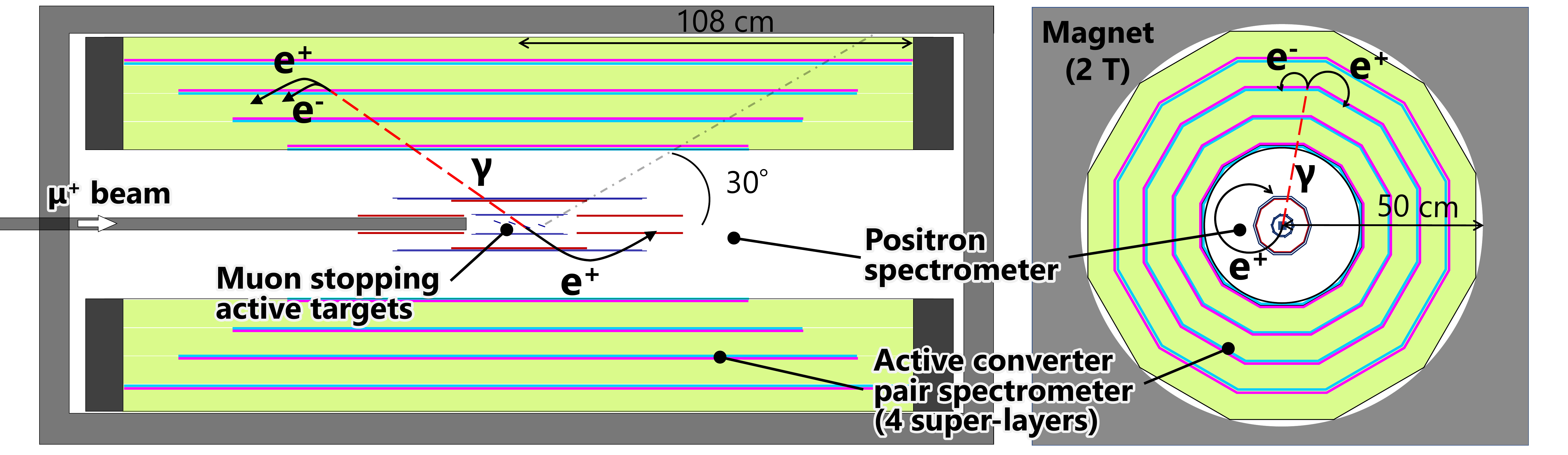}
  \caption{\textit{A sketch of a detector for \megp\ decays with a silicon pixel tracker for positrons, and active conversion layers with $e^+e^-$ trackers for photons, within a single, long
  solenoid.}}
 \label{fig:sketch}
\end{figure}

In this design, both the positron spectrometer and the photon spectrometer are housed within the same solenoid, as shown in \fref{fig:sketch}, with the solenoidal magnetic field shared by the two detectors. This configuration allows the implementation of a Mu3e-like positron spectrometer together with the photon spectrometer, while minimizing photon inefficiency arising from material placed in front of the photon spectrometer. 

The possibility of employing a gradient magnetic field, as implemented in the COBRA magnet in MEG/MEG~II will be considered 
to mitigate the saturation of the tracking system by sweeping away multiple recurling
positrons emitted at polar angles near $\theta = 90^\circ$.
The main challenge is to design a field configuration that is compatible with the requirements of both the positron and photon spectrometers.

\subsubsection{Solenoid-Toroid magnetic structure}

\begin{figure}[]
\centering
\includegraphics[width=0.8\textwidth,angle=0] 
{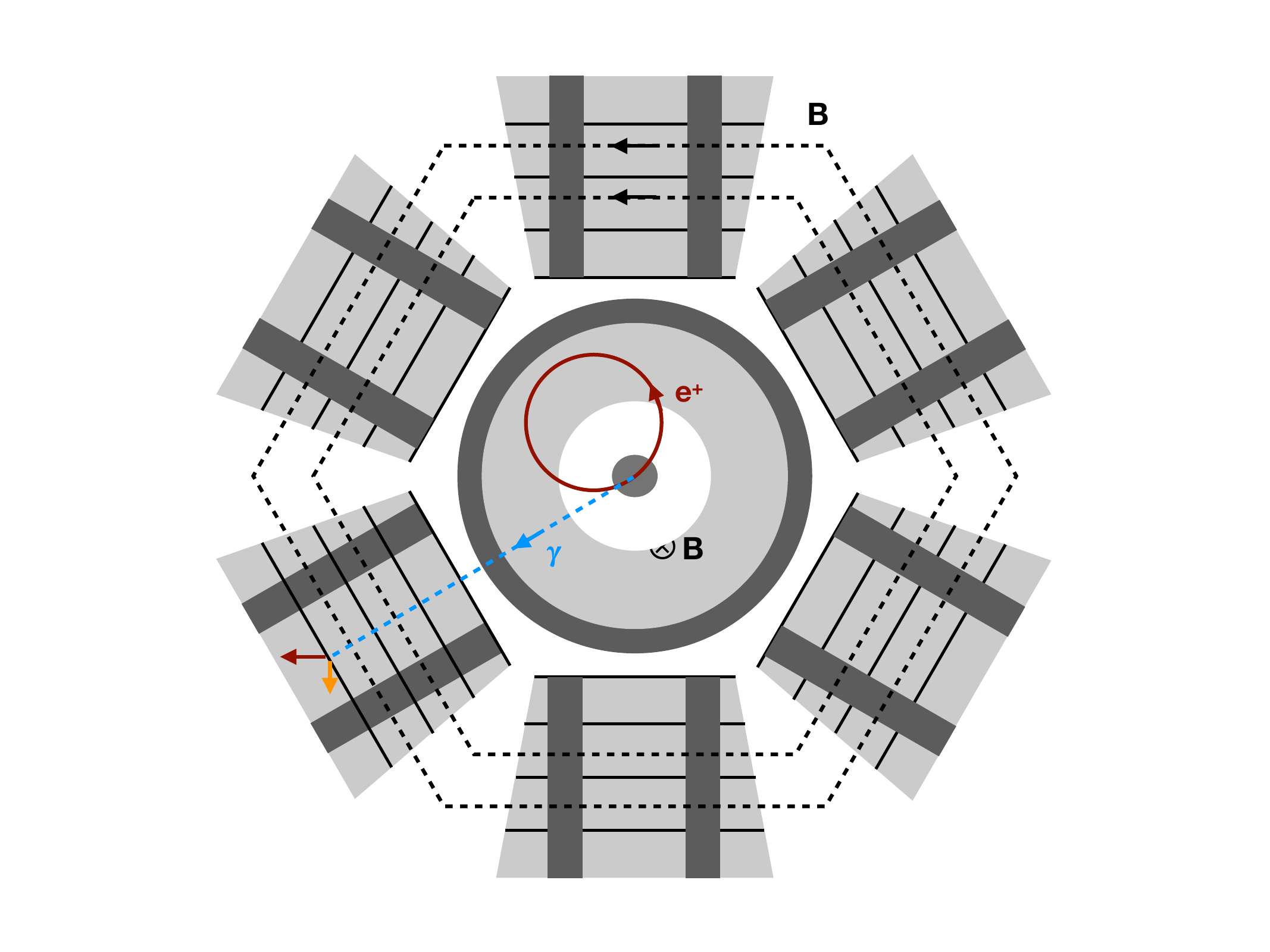}
  \caption{\textit{A sketch of a detector for \megp\ decays with a central tracker for positrons within a solenoid, and active conversion layers with $e^+e^-$ trackers for photons in an external toroidal field. 
  The hexagonal configuration is only indicative, it could be as well octagonal.}}
 \label{fig:toroid}
\end{figure}

An alternative option consists in locating the photon conversion detectors outside the solenoid, in a region where a toroidal magnetic field is generated by a set of separate coils, as sketched in \fref{fig:toroid}.

The main advantage of this geometry consists in the possibility of independently tuning the magnetic fields, in strength and shape, in the positron spectrometer and in the photon detector. 
The radial size of the photon detector can be optimized together 
with the magnetic field strength to maximize the efficiency.
In addition, one could preserve a reasonably uniform field in the gaseous tracker of the photon detector, independently of the field configuration of the inner magnet, which would make the detector calibrations less sensitive to the details of the field.

As a drawback, the material in front of the photon detector (from coils and services of both the inner magnet and the toroid) would increase the probability of photons to convert before they can be detected, resulting in a loss of efficiency. 
In MEG-II, a \SI{\sim 35}{\percent} inefficiency is ascribed to the structure of the magnet and of the LXe cryostat, despite it being specifically thinned in the relatively small polar angle acceptance region of the calorimeter.

Another possible drawback is that the particles from pair conversion are trapped within the solenoidal field and must be removed with appropriately positioned absorbers.

\section{Detectors}

\subsection{Photon reconstruction}

\subsubsection{Conversion approach}
\label{sec:conversion}
\begin{figure}[!htb]
\centering
\begin{minipage}[b]{0.49\columnwidth}
  \includegraphics[width=1.0\columnwidth] {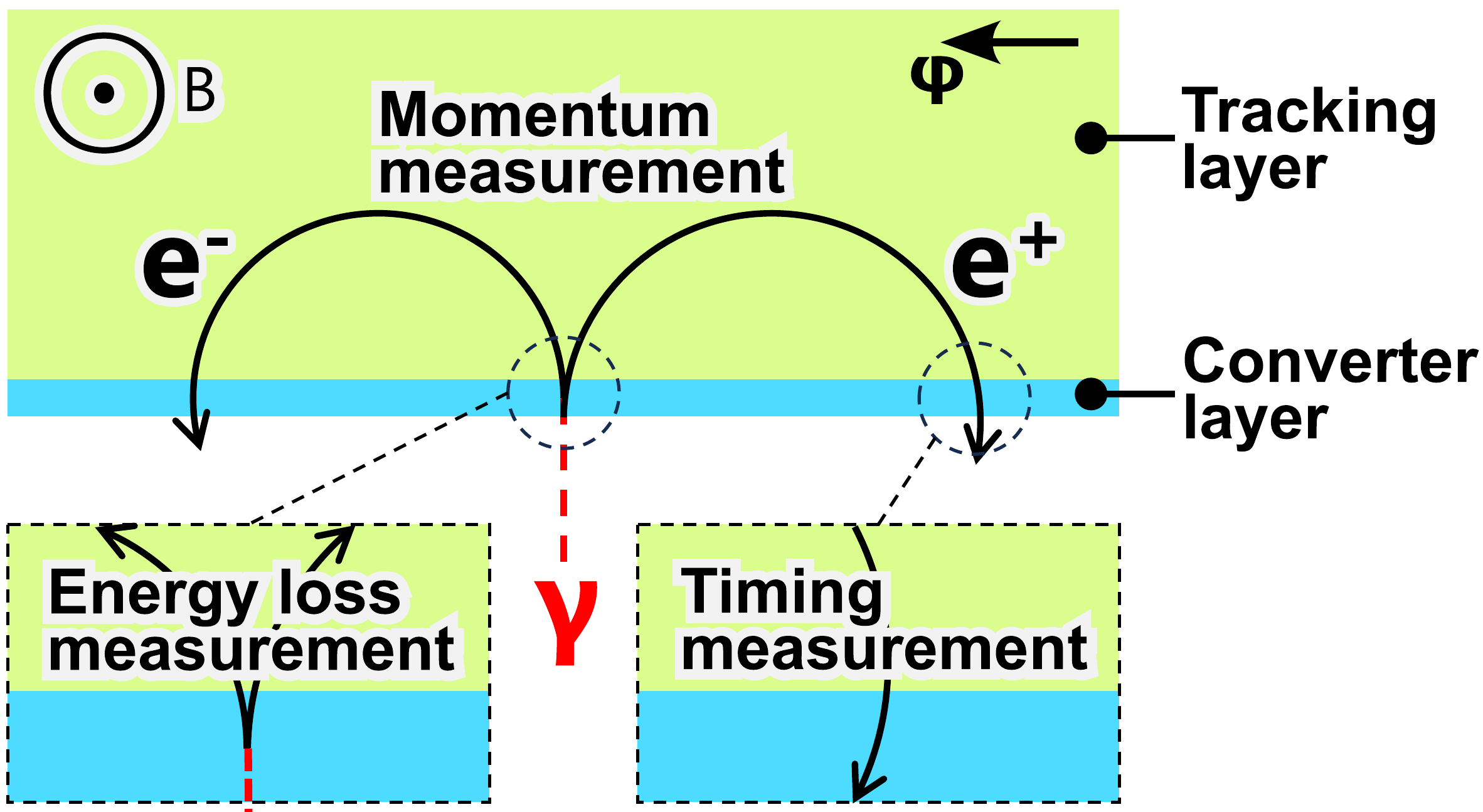}
  \caption{\textit{Concept of the active converter.}}
 \label{fig:ActiveConverterConcept}
 \end{minipage}
\begin{minipage}[b]{0.49\columnwidth}
  \includegraphics[width=0.9\columnwidth] {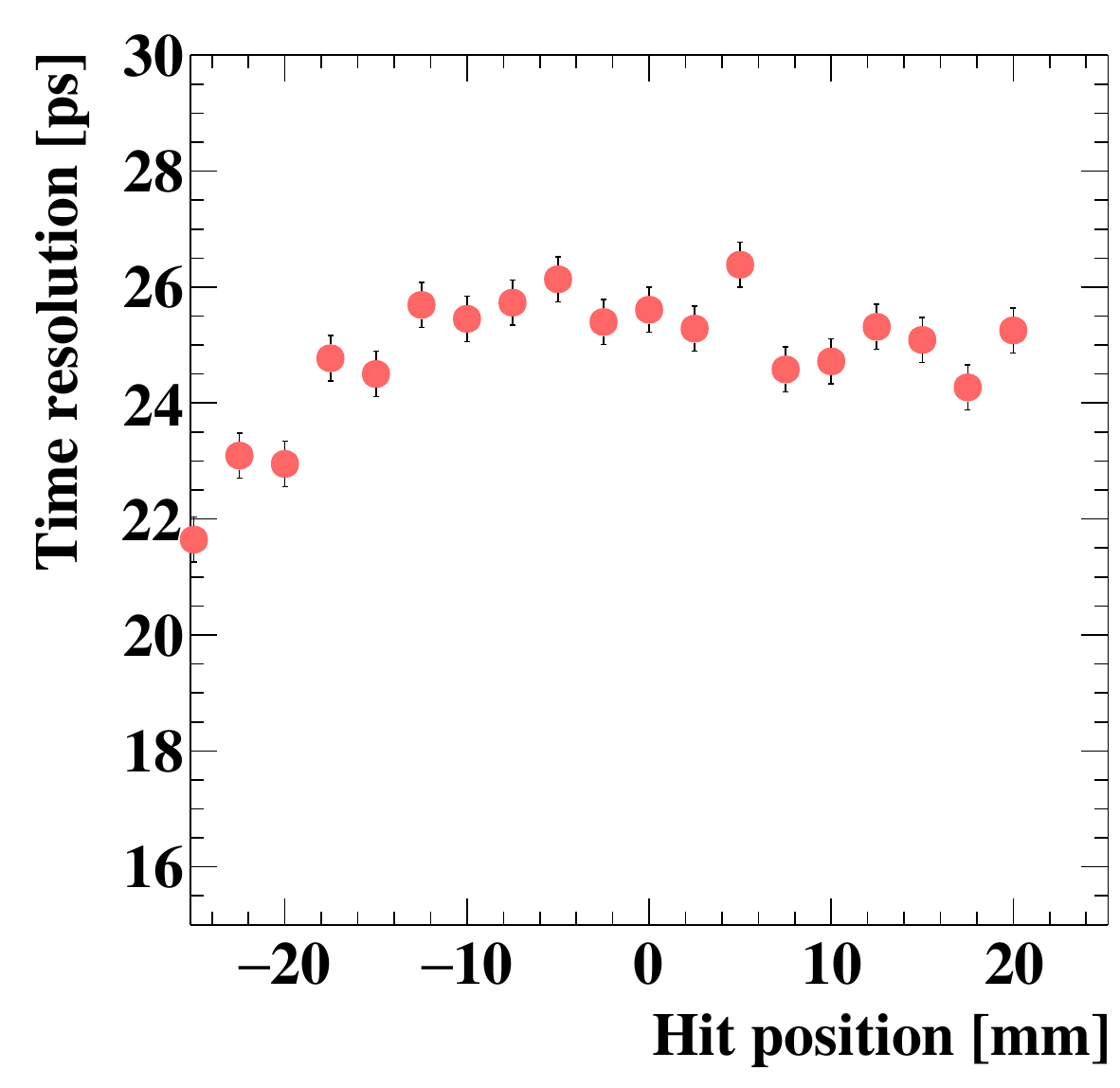}
  \caption{\textit{Single MIP time resolutions for LYSO strip of 
$5\,\mathrm{mm}\times 50\,\mathrm{mm}\times ^\mathrm{t}3\,\mathrm{mm}$.}}
 \label{fig:LYSOTimeResolution}
 \end{minipage}
\end{figure}

A conversion pair spectrometer is currently considered the baseline option for the photon 
measurement since it is more advantageous for future
$\megp$ experiments with higher beam rates. 
The photon is converted into an \positron\electron\ pair in a thin conversion layer, and the conversion 
pair is precisely tracked by a magnetic spectrometer.
The energy resolution is, therefore, expected to be much higher than the calorimetric option.
The photon hit position can also be more precisely 
measured by the reconstructed tracks.
Another advantage would be 
the capability of measuring the photon incident angle with the reconstructed tracks, which can further suppress accidental backgrounds.  
On the other hand, the limiting factors for the conversion spectrometer are the low detection efficiency and the energy loss by the conversion pair in the converter. 
The low detection efficiency can be compensated 
by multiple conversion layers and higher beam rates while suppressing the additional accidental backgrounds with higher resolutions.
For the issue of the energy loss in the converter, an active 
converter is proposed where the ionization energy loss is measured 
by the converter itself, which is then used to correct the energy measured by the conversion pair tracker (\fref{fig:ActiveConverterConcept}).
Since the energy loss in the converter is a small fraction 
of the initial photon energy, very high energy resolution is not required for the active converter. 
The active converter can also work as a photon timing detector, by
measuring the hit time of the conversion pair. 

Heavy scintillator crystals can be good candidates for the detector medium of the active 
converter with high detection efficiency.
LYSO is considered the best candidate among standard crystals thanks to its high light yield and fast response.  
A simulation study shows a detection efficiency of a few \% 
with an optimal thickness of \SI{3}{\milli\meter}.
The cylindrical conversion layer shown in \fref{fig:sketch} 
is segmented into small strips to mitigate the pileup of background photons 
and the effect of the returning conversion pair on the energy loss measurement.
The optimal segmentation is found in simulation studies to be
\qtyproduct[product-units = bracket-power]{5x50}{\mm}.
The performance of the LYSO strips was measured in the KEK AR Test Beam Line in 2023 and 2024.
\Fref{fig:LYSOTimeResolution} shows the single MIP time 
resolution of a LYSO strip of \qtyproduct[product-units = bracket-power]{5x50x3}{\mm}
read out at both ends by SiPMs (Hamamatsu MPPC S14160-6050HS
\qtyproduct[product-units = bracket-power]{6x6}{\mm}) 
measured with \SI{3}{\GeV} electrons.
Excellent time resolution of 22-27\,ps is achieved 
over the entire strip except for the edge.   
Since the photon timing is measured by both the electron and 
positron from the conversion, the time resolution is expected to be \SI{20}{\ps}, significantly improved in comparison with 
the time resolution of the MEG~II LXe photon calorimeter of 
\SI{65}{\ps}. 
The light yield of the LYSO strip was also measured to be so 
high that the contribution of the Poisson fluctuation in the number of photoelectrons is negligibly small and does not 
spoil the precise energy reconstruction of the conversion pair by the tracker.

Tracks from photon conversion will be emitted in a wide energy spectrum, and with a possibly large energy asymmetry between the \positron\ and the \electron. 
Consequently, in order to guarantee high resolution and reconstruction efficiency, extremely light pair trackers in a magnetic field are needed, covering a relatively large range of bending radii. 
Since the photon converter layers located outside the positron spectrometer region are not in a harsh environment, gaseous detectors can be used, 
providing the best possible tracking performance (thanks to the low material budget) at a relatively low cost. 
However, the limited granularity of wire chambers would make them inefficient in the low end of the energy spectrum. 
Traditional time projection chambers (TPC) are an attractive alternative, but electrons will have to drift up to \SI{1}{\meter} along the magnetic field if the solenoid option is adopted. That requires gas mixtures like Ar-CO$_2$ for a sufficiently low diffusion, which would be too heavy for this application that requires a very low material budget. For this reason, unconventional geometries need to be investigated. An interesting possibility is a design where electrons drift radially, for no more than \SI{10}{\cm}, and are multiplied in cylindrical micro-pattern gaseous detectors (MPGD). Recent advancement in cylindrical Gas Electron Multipliers (GEMs) has paved the way for this application. 
However, R\&D studies are needed for constructing cylindrical MPGDs with diameters exceeding \SI{1}{\meter}. 

For the toroidal option, the TPCs could be planar, and consequently much simpler to build, and a short drift along the azimuthal direction, that means approximately along the magnetic field, could be possible, which has the advantage to reduce the diffusion and the sensitivity to the field imperfections.

The anticipated low occupancy enables the use of stereo strips instead of pads for the readout of the TPCs, 
which helps to reduce the number of electronic channels required. 
Moreover, intriguing challenges arise in hit reconstruction and pattern recognition when using strip readout on drift distances extending over several centimeters. 
Simulations show that, with a single hit resolution around 
\SI{500}{\micro\meter}, which is a reasonable target for this kind of detectors, a resolution around $\sigma_E/E=$ \SI{0.3}{\percent}
can be reached on the 
\positron\electron\ energy sum after the energy loss in the converter. 
Hardware R\&D activities have been already initiated to study the performance of TPCs with O(\SI{10}{\cm}) drift and a strip readout.

As in MEG~II, the best resolution on the photon direction for signal events ($<\SI{10}{\milli\rad}$) would be obtained by joining the photon conversion point and the positron production point at the target. However, the precise reconstruction of the $\positron\electron$ pair would allow to reconstruct the photon direction with $\sim \SI{150}{\milli\rad}$ resolution. Having such a standalone measurement of the photon direction from the photon detector will allow to significantly reduce the accidental background, by requiring a \positron\photon\ vertex compatibility. This technique is even more effective if the muon decays are spread over a large surface. One could consider for instance to distribute the stopped muons over a few targets along the beam, staggered on the transverse plane. An optimization of the target design is therefore anticipated.

\subsubsection{Calorimeter approach}

An alternative approach based on calorimetry has also been considered as a backup solution (see \cite{Papa:2019euc, PapaLYSO, Schwendimann:2022hrz}).
Two very promising materials as BrilLanCe (Cerium doped Lanthanum Bromide, $\rm{LaBr_3(Ce)}$ and LYSO (Lutetium Yttrium OxyorthoSilicate, Lu2(1-x) Y2x SiO5 (Ce)) coupled to Silicon Photomultipliers (MPPC/SiPM) could represent an appealing option for the future calorimeter. The response of both $\rm{LaBr_3(Ce)}$ and LYSO detectors read out by SiPMs have been studied via detailed Monte Carlo (MC) simulations based on GEANT4. The impinging gammas are in the energy range 50-100 MeV. The MC simulations includes the full read out chain up to the waveform digitizer and is followed by the reconstruction algorithms.  

The results that have been obtained are very promising. 
For a cylindrical crystal (radius R = \SI{4.45}{\cm}, 
length L = \SI{20.3}{\cm}) $\rm{LaBr_3(Ce)}$ crystal the energy resolution is $\sigma_E/E =$ \SI{2.3(1)}{\percent}
and the timing resolution is $\sigma_{t}=$ \SI{35(1)}{\pico\second}.
The energy resolution can be further improved by using larger crystals (either R = \SI{6.35}{\cm} or R = \SI{7.6}{\cm}, L = \SI{20.3}{\cm}) approaching respectively $\sigma_E/E=$ \SI{1.20(3)}{\percent} or
$\sigma_E/E=$ \SI{0.91(1)}{\percent}.
Detectors based on LYSO crystals of similar size perform even better, thanks to the shorter LYSO Moliere radius compared to $\rm{LaBr_3(Ce)}$.
For a detector based on a  (R = \SI{3.5}{\cm}, L = \SI{16}{\cm}) LYSO crystal an energy resolution of 
$\sigma_E/E = \SI{1.7(1)}{\percent}$ can be obtained, 
and that can be further improved using bigger crystals 
(R = \SI{6.5}{\cm}, L = \SI{25}{\cm}, 
$\sigma_E/E = $\SI{0.74(1)}{\percent}). 
Energy resolution approaching $\sigma_E/E = $ \SI{0.3(1)}{\percent} 
can be addressed for both crystals with ultimate sizes 
(R =\qtyrange[range-phrase = --,range-units = single]{20}{23}{\cm}, 
L =\qtyrange[range-phrase = --,range-units = single]{17}{32}{\cm}),
complemented by timing and position resolutions in the range of 
$\mathcal{O}$(30) ps and $\mathcal{O}$(a few mm) respectively. 

\begin{figure}[b]
\centering
\includegraphics[width=\textwidth,angle=0]{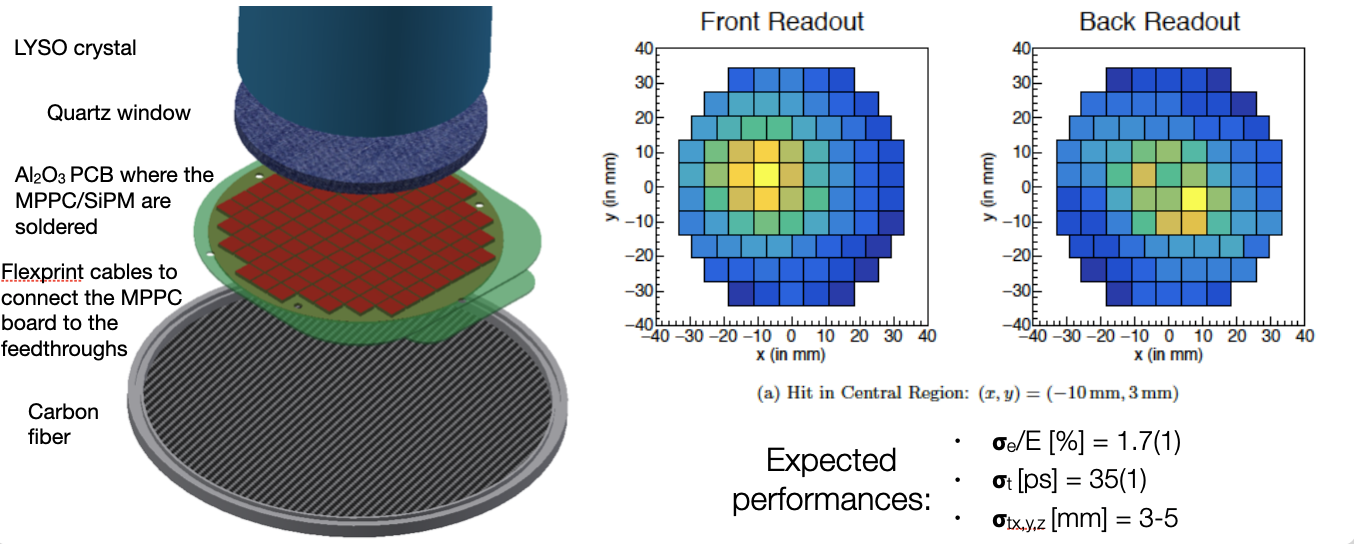}
  \caption{\textit{A sketch of a LYSO 
  crystal prototype having a \SI{4.25}{\cm} diameter and \SI{10}{\cm} length with the details of the photosensor coupling.}}
 \label{fig:LYSOprototype}
\end{figure}

Based on the availability of crystals on the market, we focused on 
building a first calorimeter prototype using LYSO crystals. 
\Fref{fig:LYSOprototype} shows a sketch of the detector, which uses a
LYSO crystal having a \SI{4.25}{\cm} diameter and a \SI{10}{\cm} length, 
equipped with a double readout scheme with SiPMs coupled to both the front and back faces of the cylindrical detector.

\begin{figure}[h]
\centering
\includegraphics[width=0.75\textwidth,angle=0]{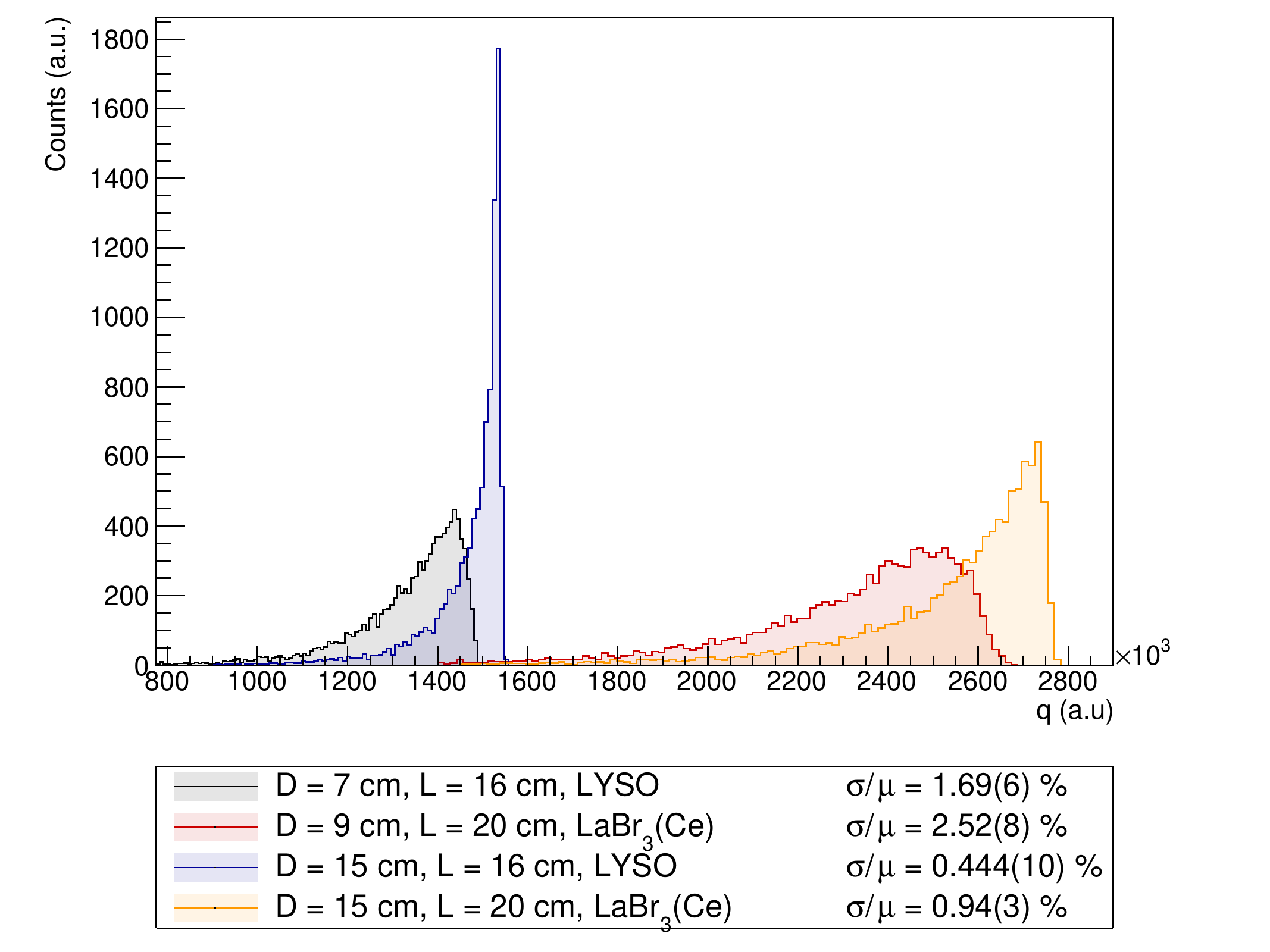}
  \caption{\textit{Expected energy resolutions for different $\rm{LaBr_3(Ce)}$ or LYSO options.}}
 \label{fig:LYSOLaBrperformances}
\end{figure}

Figure~\ref{fig:LYSOLaBrperformances} shows the ultimate resolutions that could be achieved with either a $\rm{LaBr_3(Ce)}$ or LYSO option, using the crystal sizes as reported in the caption.


\subsection{Positron reconstruction}

\subsubsection{Positron tracking}

In a scenario where the experiment is run with $R_\mu=$\SI{\sim 2e8}{\muup^+\per\second} (see Sect.\ref{sec:phase1}), the use of 
a conventional gaseous detector similar to the 
MEG-II drift chamber, possibly the chamber itself, and hence not requiring an extensive R\&D effort, could be considered for positron tracking. 
Research is currently underway on low-aging gas mixtures and innovative pattern recognition tools that use machine learning for track reconstruction. The goal is to demonstrate that such a detector can be safely and effectively operated at a beam rate more than twice that of MEG~II. 

When the muon beam rate is increased up to $R_\mu=$\SI{e9}{\muup^+\per\second} and above (see Sect.\ref{sec:phase2}), alternative options for tracking positrons need to be considered. Due to their high rate capability and granularity, silicon pixel sensors are the best choice for tracking positrons at such high muon rates.

For a positron momentum of $p=$~\SI{52.83}{\MeV/c}, from the \megp\ 2-body decay, the momentum resolution is dominated by multiple Coulomb scattering (MS). In this regime, the best momentum resolution is achieved with a tracking detector design similar to that used in the Mu3e experiment, where recurling tracks are measured after a \SI{180}{\degree} free bend. Using ultra-thin HVMAPS, which are as thin as  \SI{\sim 50}{\micro\meter}, a momentum resolution of O($100$ keV/c) can be achieved in a strong magnetic field of $B=$~\SI{2}{\tesla}. For track finding, four or preferably five tracking layers will be required to keep the rate of wrong hit combinations under control. Similar to the Mu3e pixel tracker, gaseous helium will be used for detector cooling, as it has the smallest radiation length of all known cooling technologies.

The second important handle for suppressing the accidental background is the accurate measurement of the electron-photon back-to-back topology. For this, the vertexing capability of the thin HVMAPS tracking layers, which is determined by the pixel size and the material thickness of the innermost tracking layer, and the thickness of the $\muup^+$ stopping target are of key importance. For a wide muon stopping target, which is required to fully exploit the high muon rates of HIMB, the vertex resolution will be dominated by the extrapolation uncertainty coming from MS in the innermost tracking layer. This can be remedied by an active muon stopping target, where the position of the muon decay is directly measured. Such an active target must be very thin and needs an extremely high rate capability to stand the high HIMB rates. A natural technology choice are also HVMAPS, which are highly radiation tolerant~\cite{Schoning:2020zed}. 

However, significant R\&D will be required to develop a silicon pixel detector specifically tailored for the \megp\ search.  For this reason, alternative options based on gaseous detectors are also under consideration. The advantages of a radial TPC over wire chambers could be effective once again. However, there are additional challenges posed by the extremely high occupancy levels. First, it would be essential to read the MPGDs using pads. 
This requirement, combined with a radial configuration, necessitates the development of highly compact electronics 
that can be placed on the external surface of the chamber without compromising the performance of the external detectors. 
Second, the substantial space charge that accumulates within the drift region would require advanced calibration and analysis 
tools, which have not yet been studied for the unconventional radial configuration.


\subsubsection{Positron timing}

The positron timing  must be measured to be matched with the photon timing to suppress the accidental background resulting from two 
different muons decaying close in time, one in a high energy positron and one in a high energy photon.
The timing of the positron cannot be measured close to the production point on the target because the tracking technologies under consideration do not provide an adequate resolution.
The requirement on positron timing resolution are dictated by the photon timing resolution: it must be comparable or somehow better.
From \sref{sec:conversion}, where we assume a photon timing resolution less than 30~ps, the design goal of the positron timing 
detector is \SI{\sim 20}{\pico\second}. We also have to guarantee that the correction for the positron time of flight to the decay target adds a small contribution to the resolution.

The resolution of the pTC in MEG~II \SI{\sim 30}{\pico\second} \cite{megII-det} is already close to the requirements. That is obtained relying on plastic scintillator pixels readout on opposite sides by array of SiPMs.
They are arranged in such a way that several ($\sim$ 9) are traversed by
signal positrons.
A resolution improvement can be achieved by reducing the size of the 
pixels, increasing their thickness, the number of SiPM, the SiPM size.
Alternatively, the design of Mu3e can be considered, where one or two cylindrical layers of small scintillating cubes (\qtyproduct[product-units = bracket-power]{\sim 1x1x1}{\cm}) 
are readout by a single SiPM; again the signal positron is expected to cross multiple pixels.
A focused R\&D program with tests and simulations will determine the optimal balance between timing performance, cost, mechanical complexity 
and number of readout channels. 

For the future (see Sect.\ref{sec:phase2}),
alternative solutions for positron timing are also under study, thanks to the recent developments of fast timing detectors.
The HVMAPS themselves could measure the positron time with a very good resolution. 
One of the HVMAPS tracking layers could also serve as a dedicated timing layer, providing a time resolution $\lesssim$ \SI{50}{\pico\second} per particle crossing, which have already been demonstrated with HVMAPS prototypes using two different approaches, firstly by using fast bipolar transistors for signal amplification~\cite{Martinelli:2021cbl} and secondly by using a pre-processed wafer with an internal gain layer~\cite{Milanesio:2022vjk}. 
Additionally, we are considering options based on low-gain avalanche detectors (LGAD)~\cite{Croci:2023fzs} or gaseous detectors with extreme time resolutions (PICOSEC)~\cite{Bortfeldt:2017daz}.

\section{Experimental roadmap}
\label{sec:schedule}

To optimize the use of time and resources, we are considering to adopt a staged approach, with a phase-I experiment with limited performance but still improving significantly the MEG~II sensitivity, followed by a phase-II experiment to reach the final sensitivity.

The core of the experimental proposal being the converter as photon detector (see \ref{sec:conversion}), these stages need to be preceded by an experimental proof of the efficiency and resolutions that can be obtained with this approach (phase-0). 

\subsection{Proof-of-concept of the conversion technique (Phase-0)}

The converter consists of high density scintillating crystals for timing and energy measurements and of a tracker, possibly a radial TPC, for momentum and direction measurement.
It requires an extensive set of simulations and laboratory and beam tests to choose the most effective detection technologies for the crystals and the tracker as 
well as defining the geometry and the magnetic field requirements.

\begin{figure}[htb]
    \centering
    \subfloat[][Sketch of a setup with a dipole magnet for testing a pair converter.]{\includegraphics[width=0.45\textwidth]{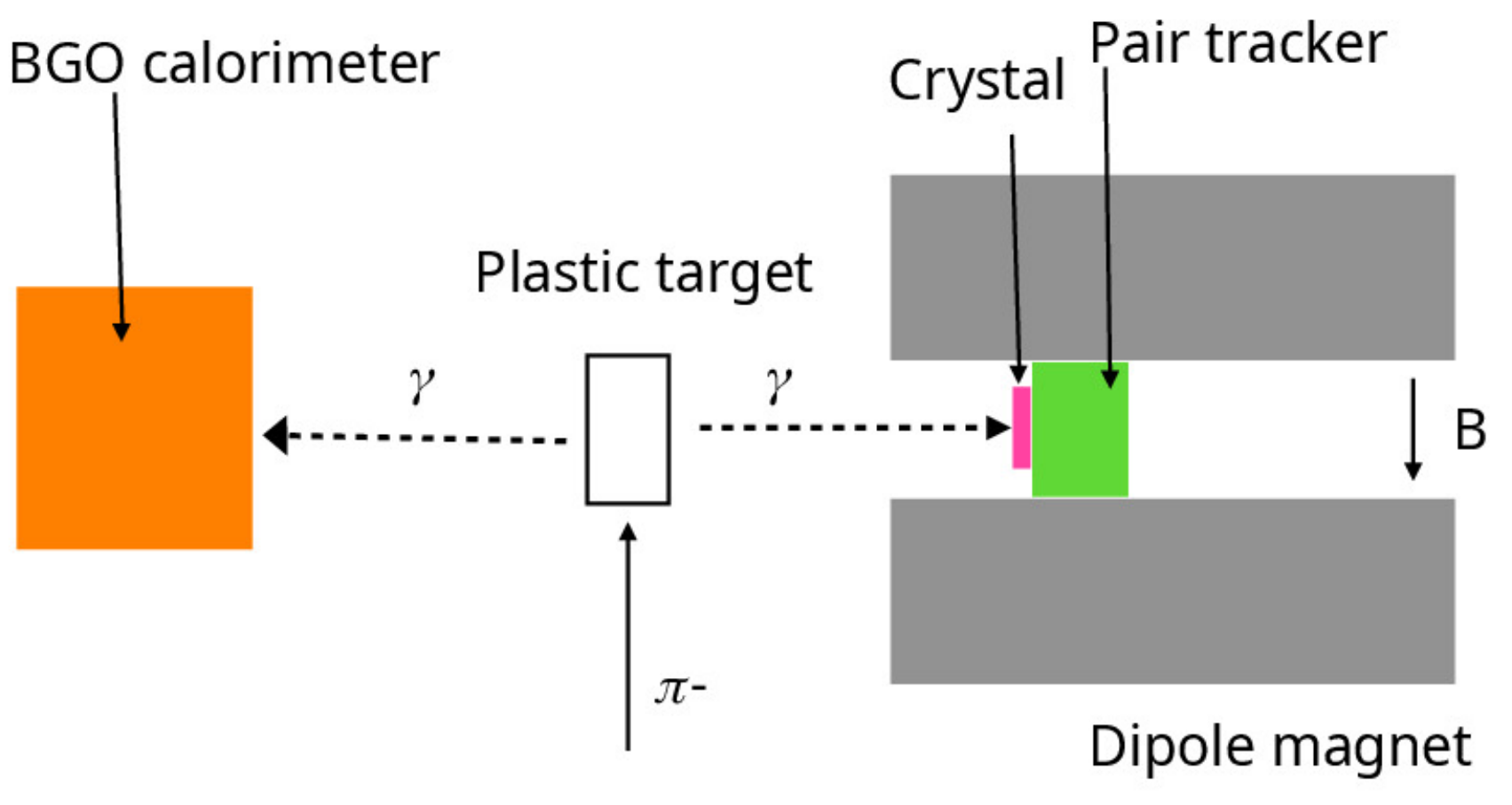}}
    \hspace{0.75cm}
    \subfloat[][Sketch of a setup with the COBRA magnet for testing a pair converter.]{\label{fig:fig2}\includegraphics[width=0.45\textwidth]{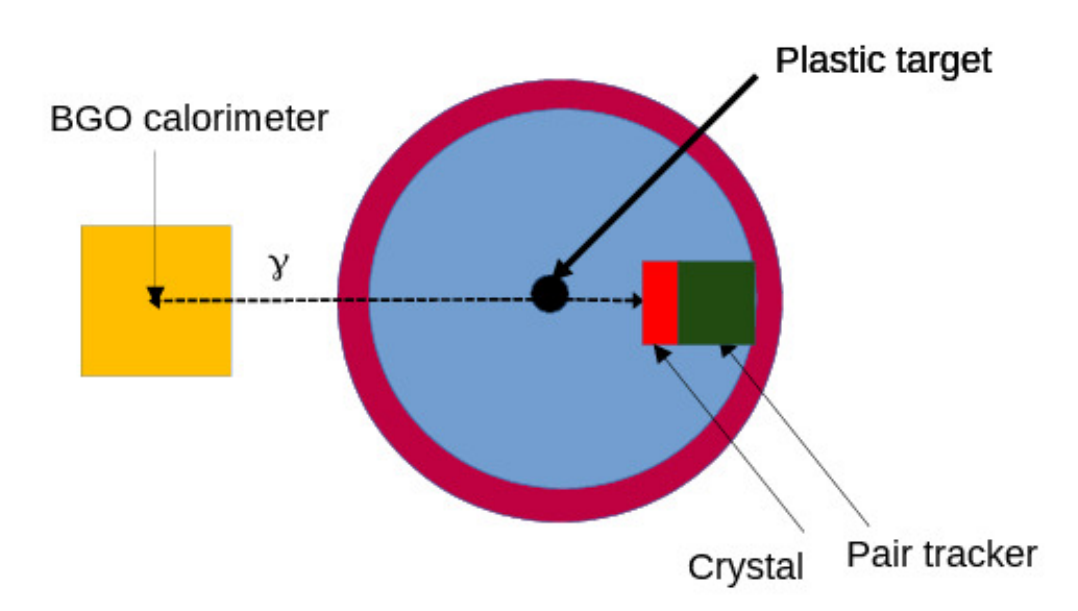}}
  \vspace{0.5cm}
  \caption{\label{fig:Phase0}Possible setups for a proof-of-concept experiment for the photon conversion technique.}
\end{figure}

Some beam tests with electrons impinging on Ce-doped LYSO crystals have been recently performed
\cite{Sakakibara:2025yhk,Ban:2025sev} providing much needed information on time resolution and light yield.
Other beam tests are required with photons, 
the crystal embedded in a strong magnetic field 
followed by a TPC for measuring the momenta of the
$\positron\electron$ pairs.
Ideally, those tests require a tagged photon beam 
with known energy $\sim$\SI{50}{MeV}. A possible 
approach is to use a $\pi^-$ beam impinging a plastic target exploiting the Charge Exchange Reaction (CEX) $\pi^-p\to n \pi^0$ followed by 
$\pi^0\to \gamma \gamma$. As sketched in 
Fig.\ref{fig:Phase0} one photon will be tagged by 
a BGO calorimeter, that will measure its energy,
and the other will interact with the crystal followed by a small TPC. 
In the option on the left the converter prototype will be embedded in a dipolar field, 
in the option on the right, the magnetic field will be provided by the COBRA magnet. 
Other configurations are possible and the tests will be planned taking into account the PSI long shut down and the availability of other accelerator complex. 
Several tests will be required to define the design parameters of the converter including the readout of the crystals and the TPC, the thickness and number of the converter crystals, the intensity of the magnetic field, the TPC mechanical and gas properties.


Considering the R\&D focus in Phase-0 of this proposal, the TDAQ system employed has to be easily adaptable to various scenarios. 
For this reason, we envisage the reuse of MEG II TDAQ based on the DRS4 chip \cite{francesconi2023wavedaq}:
saving the complete waveform will let us investigate the ultimate resolutions of the technology under study, that is 
an important input to select the most appropriate analog frontend technology for the final detectors. 
We are aware that the TPC signals will not fit in MEG II TDAQ constraints because of the long drift time. We already have experience in pairing the MEG II TDAQ to external DAQ systems, which, for this particular technology, could be based on CERN SRS system \cite{de2023operation}.

\subsection{Phase-I}
\label{sec:phase1}

The success of the proof-of-concept experiment will pave the way toward an experiment to be performed at HIMB to reach the ultimate sensitivity goal of this project. 
However, in consideration of the R\&D and engineering activities which will be necessary to reach this stage, and the Mu3e operations at HIMB in the early 2030s, an experiment fully satisfying the requirements for O($10^{-15}$) sensitivity will only be possible in the second half of the next decade. 

For these reasons, we are considering operating a phase-I apparatus, to be operated in the early 2030s on a conventional beam line. 
It would be a version including several components 
of the final experiment, possibly with a reduced number of conversion layers, to reduce the costs and simplify the engineering of the detector. 
The goal is twofold: 
\begin{enumerate}
    \item Improve significantly the sensitivity to \meg\ with respect to MEG~II
    \item Develop experimental solutions for the final phase-II apparatus:
    \begin{itemize}
        \item Magnetic structure
        \item \positron\electron converter
        \item Trigger and readout at high rate
    \end{itemize}
\end{enumerate}.
The improvement in sensitivity in a relatively short time-scale opens the road to a possible discovery and offers a strong motivation to the collaboration to push forward with the project.

Since the development of a silicon pixel tracker will require the prior completion of the Mu3e program, running the phase-I experiment on a conventional beam line (namely $\pi$E5) with up to \SI{2e8}{\upmu^+/\second}, would make possible using a conventional gaseous detector as the MEG~II drift chamber for positron tracking. It would also avoid any conflict with the beam-usage schedule of Mu3e Phase-II, which will run at HIMB in the same period.

In the scenario with a single solenoid hosting both the positron and photon detectors, a new solenoid magnet, possibly with a graded magnetic field, needs to be designed and built in the next years: 
the same magnet is likely be reused in phase-II. If instead we will opt for a solenoid-toroid configuration, the MEG~II magnet COBRA could be used for positron tracking in this phase, while a new solenoidal magnet will be designed and built for phase-II profiting from the experience gained on phase-I. 
In both cases, if the adopted solution involves a gradient magnetic field as in MEG~II, a beam transport magnet is necessary to couple the detector solenoid to the beam line. The MEG~II Beam Transport Solenoid (BTS) cannot be reused, because it requires continuous supply from a liquid helium line that is going to be dismantled in 2027. Consequently, a new beam transport magnet would have to be designed and built already for phase-I.

The phase-I, and subsequent phase-II, experiments will require a complete rethinking of the TDAQ topology, in particular for the trigger section: in MEG II, the $E_\gamma$ trigger condition is crucial in reducing the trigger rate. 
However, for the converter based design, the $E_\gamma$ information is encoded in the pair kinematics and therefore needs the full 3D track reconstruction already at the trigger level. 
This scheme is already possible with current streaming DAQ technologies 
but will require a considerable R\&D effort to optimize the hit streaming from detectors and the triggering reconstruction.

The main challenges of phase-I are:
\begin{enumerate}
\item the design and production of the magnets;
\item the optimization of the target and detector geometry;
\item the optimization and the engineering of the active conversion layers;
\item the selection, optimization and test of the technology for the readout of the $\positron\electron$ tracker among the existing MPGDs, with particular care to the mechanical aspects if large cylindrical detectors are needed;
\item the demonstration of the operability of a drift chamber at $R_\mu \sim \SI{2e8}{\muup^+\per\second}$, in terms of both hardware stability and track reconstruction;
\item the development of high-granularity positron timing detectors;
\item the realization of an appropriate trigger and DAQ architecture to handle the significantly higher number of readout channels with respect to MEG~II and the possible variety of acquisition platforms.
\end{enumerate}

The information that phase-I will give on the subsystems is extremely important and will allow a more robust design of the phase-II experiment.
It would prevent the risks of approaching the ultimate experiment in a single step, with a rate increase of a factor of 20, a completely novel photon converter, an untested magnetic structure and with very demanding trigger and readout conditions.

\subsection{Phase-II}
\label{sec:phase2}

The phase-II will profit of the experience gained in phase-I, from Mu3e and from R\&D in detector technologies like radial TPC at high rate. Based on this knowledge, a robust detector design will be produced to
match the challenge at $R_\mu \geq \SI{1e9}{{\muup^+\per\second}}$.

The most critical task will be the development of a positron tracker capable of tolerating the high fluences that such a muon beam rate implies. We will take advantage of the experience gained by the Mu3e collaboration in the meanwhile. For this reason, the construction of the apparatus would follow the completion of the Mu3e experimental program, while minor upgrades of the other detectors could be performed in parallel. The design and construction of a dedicated solenoid will also be necessary if not already done for phase-I.

Based on the experience accumulated in phase-I, it will be also possible to perform an incremental upgrade of the photon detector and the positron timing detector, going beyond a mere increment of the number of conversion layers or an extension of the acceptance. New technologies for positron timing could be further considered at this stage.

The phase-II \megp\ experiment, incorporating the new tracker and the multi-layer converter and operating with dedicated magnets, would be performed at HIMB in the second half of the next decade.

\subsection{Schedule}
 
\begin{figure}[t]
\centering
\includegraphics[width=\textwidth,angle=0] {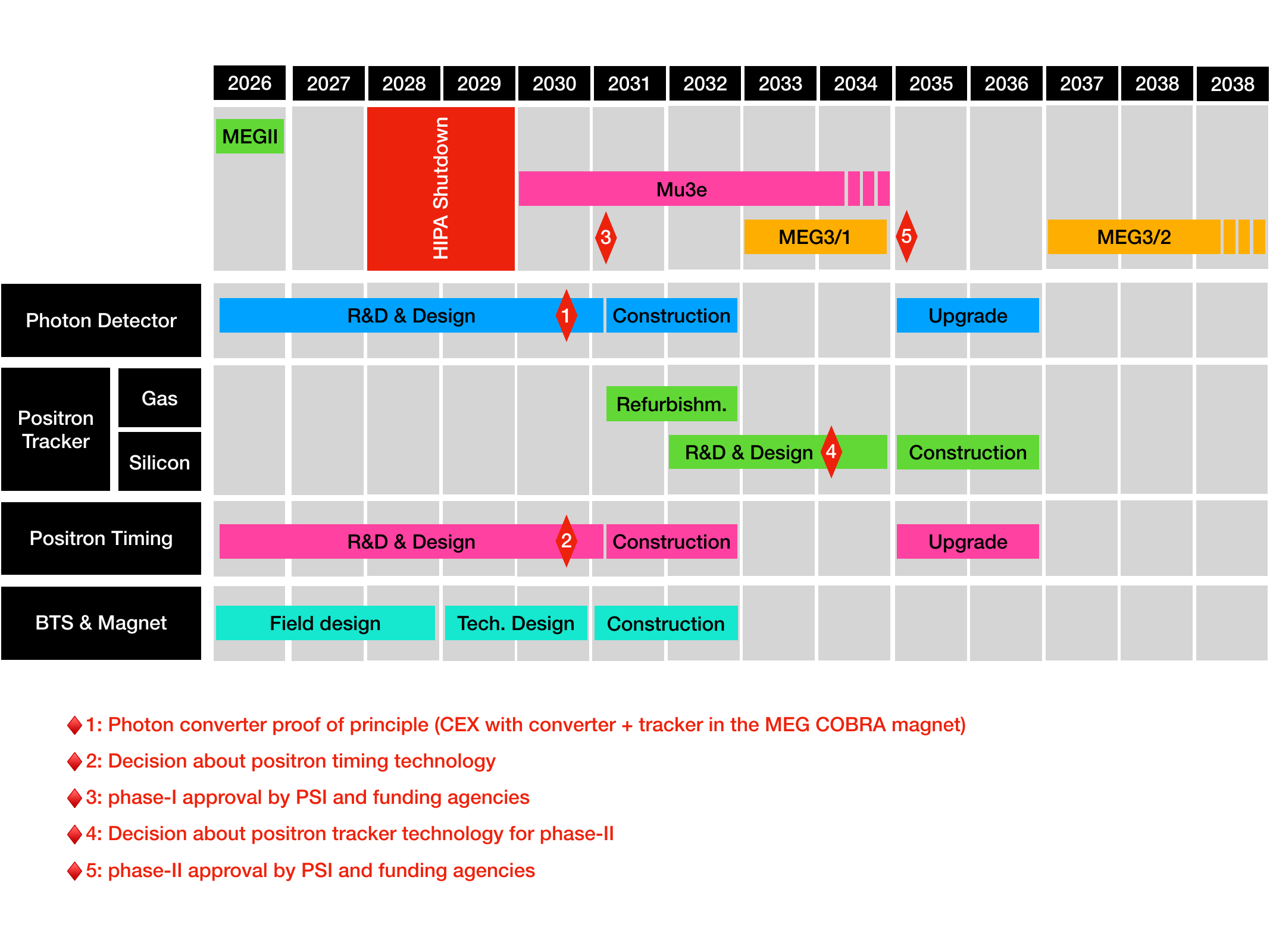}

  \caption{\textit{A tentative schedule of the experimental program for a future \megp\ experiment in two phases (MEG3/1 and MEG3/2).}}
 \label{fig:schedule}
\end{figure}

\Fref{fig:schedule} shows a tentative schedule, considering the option of adopting a single dedicated solenoid for both phase-I and phase-II.

\section{Sensitivity}

\subsection{Detector performance}

Table\ref{tab:detector-performance} summarizes the detector performance used in the sensitivity estimates. The values for the positron tracking are inferred from the MEG-II data, the values for the converter are estimated from simulations. The photon reconstruction efficiency ($\sim\SI{3.3}{\percent}$ per layer), which is also estimated from simulations, is a combination of the photon conversion efficiency ($\sim\SI{12}{\percent}$ at \SI{52.8}{\MeV/c}), the probability of having both the electron and the positron with a transverse momentum large enough to be tracked ($\sim\SI{60}{\percent}$), the efficiency for reconstructing and fitting them ($\sim\SI{80}{\percent}$), and the probability of getting a reconstructed energy within a likelihood analysis region of $\pm 3\sigma_{E_\photon}$, when considering radiation losses and other effects contributing to the tails of the  energy resolution ($\sim\SI{57}{\percent}$).
For phase-I, if the COBRA magnet of MEG-II is used in combination with a toroid, the geometrical acceptance for the photon is conservatively limited to the central region of the magnet, corresponding to the polar angle acceptance of the LXe calorimeter, because only in this region the magnet was designed with minimal material to guarantee a high photon transmission. The resulting acceptance is \SI{36}{\percent}.

For phase-II, five separate stopping targets are assumed 
to further reduce the accidental backgrounds using the 
photon angle measured at the pair spectrometer, for phase-I this variable has a more limited discriminating power. Conservatively, we have assumed no discriminating power in phase-I.

\begin{table}[!htb]
 \centering
 \caption{Summary of the detector resolutions and efficiencies for four layers' converter. 
 It is assumed that the tracking performance is the same for Phase-I and Phase-II even if the detectors are expected to be different. The geometrical acceptance in parenthesis is for phase-I utilizing the MEG~II COBRA magnet.}
 \vspace{1em}
 \label{tab:detector-performance}
 \begin{tabular}{@{}lc}
 \hline
 & Resolutions/efficiencies \\
 \hline\hline\noalign{\smallskip}
 Photon energy & \SI{200}{keV} \\
 Photon position & \SI{200}{\micro m} \\
 Photon timing & \SI{30}{ps} \\
 Photon angle & \SI{150}{mrad} \\
 Photon detection efficiency (4 layers) & \SI{10}{\percent}\\
 Positron energy & \SI{100}{keV} \\
 Positron angle & \SI{6}{mrad} \\
 Positron timing & \SI{30}{ps} \\
 Positron detection efficiency & \SI{70}{\percent} \\
 Geometrical acceptance & \SI{85}{\percent} (\SI{36}{\percent}) \\
 \hline
 \end{tabular}
\end{table}

\subsection{Sensitivity estimate}
\label{sec:senscalc}

\begin{figure}[t]
        \centering
        \includegraphics[width = 0.5\textwidth]{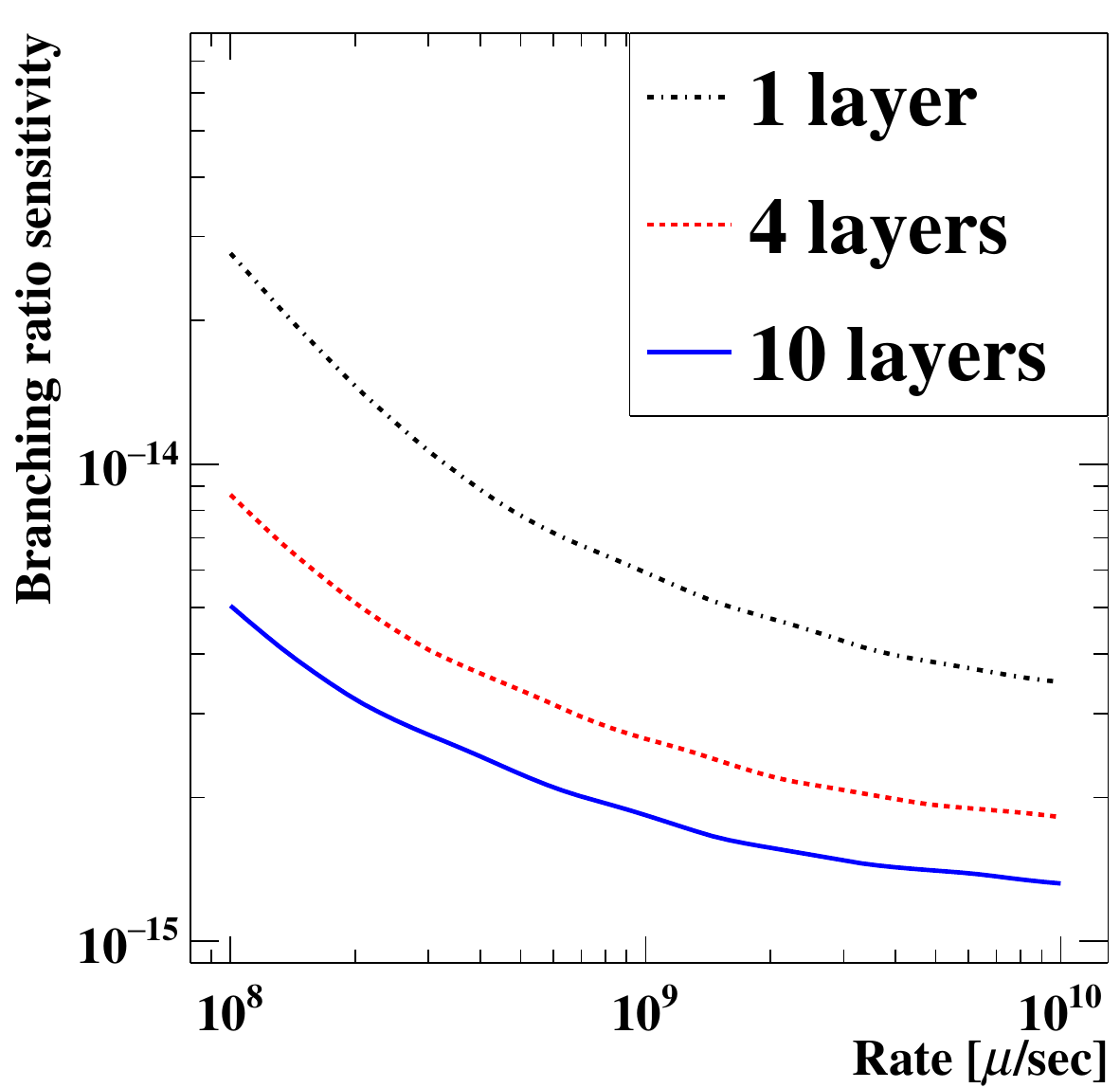}
        \caption{\textit{
        The projection of the branching ratio sensitivity 
        (90\% C.L. upper limit) for a 3-year run 
       as a function of the beam rate
        for the experiment with the conversion pair spectrometer.
        Three scenarios for the number of conversion layers are compared.
        }}
	\label{Sensitivity}
\end{figure}

An estimate of the projected sensitivity was made, based on a likelihood analysis that is a simplified version of the one of MEG~II. \Fref{Sensitivity} shows the projected sensitivity for the scenario with single solenoidal magnet (large geometrical acceptance) versus the beam rate. Three scenarios for the number of conversion layers are compared; one layer, four layers as the baseline and ten layers as an extreme case.
A sensitivity around \num{1.5e{-14}} is obtained with a beam rate of \SI{2e8}{\muup^+\per\second} in phase-I with a single conversion layer, already much better than the MEG~II final sensitivity. The sensitivity goes down to $(2-3)\times 10^{-15}$ in phase-II at a beam rate above \SI{e9}{\muup^+\per\second} with four conversion layers. 

If the MEG-II COBRA magnet is used in phase-I, with the converter in the toroid, the smaller geometrical acceptance worsens the sensitivity, requiring two conversion layers to keep the sensitivity around \num{1.5e{-14}}.

Referring to \fref{fig:theory}, for $|\kappa_D| \leq 1$ the expected NP scale sensitivity of phase-I is better than the expected final Mu3e result in the \mutec\ channel and only slightly worse than the expected Mu2e final result in the \convN\ channel; this result will be bettered in phase-II leaving only AMF/PRISM to potentially provide a better NP scale sensitivity.

\clearpage

\bibliographystyle{my}
\bibliography{MEG}
\end{document}